At the end of this tex-file we put the 4 figures (that belong to this
manuscript), compressed with "uufiles". Read there how to unpack this
file. 
You can print the tex-file and the 4 ps-files (figures) separately, but
using "dvips" you get the figures printed at the right place in the text.
----------------------------------------------------------------
\documentstyle[12pt,a4]{article}   
\newcommand{\etal}{\mbox{\it et al.\/}}
\newcommand{\eg}{\mbox{\it e.g.\/}}

\begin{document}

\title{{\LARGE $\bar{p}p$ partial-wave analysis and \\
       $\overline{N}\!N$ potentials}}
\author{{\it J.J.\ de Swart and Th.A. Rijken} \\
        Institute for Theoretical Physics,
        University of Nijmegen  \\
        Nijmegen, The Netherlands  \\[2mm]
         and \\[2mm]
        {\it R. Timmermans} \\
        Institute for Theoretical Physics,
        University of Nijmegen  \\
        Nijmegen, The Netherlands \\  and \\
        Theoretical Division,
        Los Alamos National Laboratory \\
        Los Alamos, NM 87545, USA}
\date\today
\maketitle

\begin{abstract}
A review is given of the different ways to describe $\bar{p}p$ scattering.
Next the Nijmegen partial-wave analyses of the $\bar{p}p$ data as well as
the corresponding Nijmegen $\overline{N}\!N$ database are discussed.
These partial-wave analyses are finally used as a tool to construct
a better $\overline{N}\!N$ potential model and also to clarify questions
raised in the literature.
\end{abstract}

\vfill

\begin{center}
Invited talk given by J.J. de Swart at the Second Biennial Workshop on \\
Nucleon-Antinucleon Physics, NAN '93, Moscow, Russia, September 1993.
\end{center}

\newpage
\section{Introduction}         \label{I}
Starting with an antiproton beam directed on a proton target many reactions
are possible. First of all is the elastic scattering $\bar{p}p \rightarrow
\bar{p}p$. For this reaction differential cross sections $\sigma_{el}(\theta)$,
analyzing-power data $A_{el}(\theta)$~\cite{Kun88,Ber89}, and even
some depolarization data $D_{yy}(\theta)$~\cite{Kun91} have been
measured. We will discuss these extensively in this talk.
The annihilation channel, $\bar{p}p \rightarrow$ mesons, is studied very 
intensively by theorists as well as by experimentalists. Many different
reactions can be distinguished. For our purposes, however, only a global
description will turn out to be sufficient. The charge-exchange
reaction $\bar{p}p \rightarrow \bar{n}n$ has its threshold
at $p_L = 99$ MeV/c. Important is that in the one-boson-exchange (OBE)
picture only charged mesons can be exchanged. The most important of these
are the $\pi^{\pm}$ and $\rho^{\pm}$ mesons. The study of this reaction
allowed us to determine the coupling constant of the charged pion
to the nucleons~\cite{Tim91b}. Recently, excellent data for
this charge-exchange reaction has been obtained at LEAR for
the differential cross section $\sigma_{ce}(\theta)$
and for the analyzing power $A_{ce}(\theta)$~\cite{Bir90}.
Very recently, even charge-exchange depolarization data have
become available~\cite{Bir93}.
Excellent data are also available for the strangeness-exchange 
reaction $\bar{p}p \rightarrow \bar{\Lambda}\Lambda$
with threshold $p_L = 1.435$ GeV/c~\cite{Bar87}.
More data for this reaction are forthcoming as well
as data for the other strangeness-exchange reactions 
$\bar{p}p \rightarrow \bar{\Lambda}\Sigma,
\bar{\Sigma}\Lambda$~\cite{Bar90} and $\bar{\Sigma}\Sigma$.
These reactions are very important for the precise determination
of the $\Lambda N\!K$ and $\Sigma N\!K$ coupling constants
and combining these with the $N\!N\pi$ coupling constant
gives us information about flavor SU(3)~\cite{Tim91a}.

\section{Antinucleon-nucleon potentials}         \label{II}
It is customary to start with some meson-theoretic $N\!N$ potential
and then apply the $G$-parity transformation \cite{Pai52}
to get the corresponding 
$\overline{N}\!N$ potential. This is a straightforward, but rather
cumbersome procedure. When you ask people about details,
then most people must confess that they do not know.

We would like to point out that just charge conjugation,
together with charge independence, without actually combining
them to $G$, is sufficient for our purposes.
To understand this, let us look at the $ppm^{0}$ vertex describing
the coupling of a neutral meson $m^{0}$ to the proton $p$ with a
coupling constant $g$.
When we apply charge conjugation $C$ we have 
\[
\bar{p} = Cp \rule{1cm}{0mm} {\rm and}
 \rule{1cm}{0mm} \overline{m^{0}} = Cm^{0}\ ,
\]
and we describe now the $\bar{p}\bar{p}\overline{m^{0}}$ vertex.
For nonstrange, neutral mesons $m^{0}$ one can define the charge 
parity $\eta_{c}$ by $\overline{m^{0}}=\eta_{c} m^{0}$. 
Charge-conjugation invariance of the interaction
Lagrangian describing this $ppm^{0}$ vertex requires that the
coupling constant $g$ of the meson $m^{0}$ to the proton $p$ is equal to
the coupling constant of the antimeson $\overline{m^{0}}$ to the antiproton
$\bar{p}$. The coupling constant $\bar{g}$ of the meson $m^{0}$
to the antiproton is then given by
\[
\bar{g} = \eta_{c} g\ .
\]
For mesons of the type $Q\overline{Q}$, with relative orbital
angular momentum $L$ and total spin $S$ the charge parity is
\[
\eta_{c} = (-)^{L+S}\ .
\]
Therefore the pseudoscalar ($^{1}S_{0}$) mesons have $J^{PC} = 0^{-+}$,
the vector ($^{3}S_{1}$) mesons have $J^{PC} = 1^{--}$,
the scalar ($^{3}P_{0}$) mesons have $J^{PC} = 0^{++}$, etc.

We see that from the important mesons only the vector mesons have
negative charge parity and therefore the coupling constants of the vector
mesons change sign when going from the nucleons to the antinucleons.
In the OBE picture the $pp$ potential $V(pp)$ is the sum of the exchanges
of the pseudoscalar meson $\pi$, the vector mesons $\rho$ and $\omega$,
the scalar meson $\varepsilon(760)$, etc. That is
\[
V(pp) = V_{\pi} + V_{\rho} + V_{\omega} + V_{\varepsilon} + \ldots
\]
The potential $V(\bar{p}p)$ described in the same OBE picture is
then given by
\[
V(\bar{p}p) = V_{\pi} - V_{\rho} - V_{\omega} + V_{\varepsilon} + \ldots
\]
In these reactions only neutral mesons are exchanged. When we want to
describe the charge-exchange reaction $\bar{p}p \rightarrow \bar{n}n$, then
it is easiest to recall charge independence. Charge independence requires
that the coupling constant $g_{c}$ of the charged meson to the nucleons
is given by $g_{c} = g\sqrt{2}$, and to the antinucleons by
$\bar{g}_{c} = \bar{g}\sqrt{2}$. The charge-exchange potential
is therefore given by
\[
V_{ce} = 2(V_{\pi} - V_{\rho} + \ldots)\ .
\]
The diagonal potential in the $\bar{n}n$ channel is, using
charge independence, given by $V(\bar{n}n) = V(\bar{p}p)$.

{\it What can we learn from the $N\!N$ potentials about the 
$\overline{N}\!N$ potentials}? \\
The $pp$ central force is relatively weak due to the cancellations between
the repulsive contribution of the vector mesons $\omega$ and $\rho$ and
the attractive contribution of the scalar mesons $\varepsilon(760)$, etc.
The $\bar{p}p$ central force is strongly attractive, because the
vectors mesons have now an attractive contribution which adds 
coherently to the attractive contribution of the scalar mesons,
giving a very strong overall central force. 
Also the tensor force in $N\!N$ is relatively weak, because
$\pi$ and the important $\rho$ contributions have opposite sign.
In the $\overline{N}\!N$ case these mesons add coherently again to
give a very strong tensor force. This strong tensor force is
responsible for the importance of the transitions
\[
\ ^3S_1 \leftrightarrow\ ^3D_1 \rule{1cm}{0mm} , \rule{1cm}{0mm}
\ ^3P_2 \leftrightarrow\ ^3F_2 \rule{1cm}{0mm} , \rule{1cm}{0mm}
\ ^3D_3 \leftrightarrow\ ^3G_3 \rule{1cm}{0mm} {\rm etc.}
\]

\section{Various models}              \label{III}
\subsection{Black-disk model}         \label{III.1}
One of the simplest models for the description of the elastic and
inelastic cross section is the black-disk model. This model gives
\[
\sigma_{el} = \sigma_{ann} = \frac{1}{2} \sigma_{T} = \pi R^{2}\ ,
\]
where $R$ is the radius of the black disk.
This relation is satisfied very approximately. It shows that the 
annihilation cross section predicts radii for the black disk
which are energy dependent and pretty large. In the 
momentum interval 200 MeV/c $< p_L <$ 1 GeV/c
this radius $R$ varies from more than 2 fm to about 1.4 fm.

\subsection{Boundary-condition model} \label{III.2}
The boundary-condition model in $\overline{N}\!N$ was first
introduced by M. Spergel in 1967~\cite{Spe67}. Later many
more people used this now more than a quarter century old
model (see \eg\ Refs.~\cite{Dal77,Del78}).
The model is based on the observation that the interaction for large
values of the radius is often well-known, while the interaction for small
radii is very hard to describe. This problem is then solved by just
specifying a boundary condition at $r=b$. For this boundary condition 
one takes the logarithmic derivative of the radial wave function at the
boundary radius $b$
\[
P=b \left( \frac{d\psi}{dr}/\psi \right)_{r=b}\ .
\]
Outside this radius one assumes that the interaction can be described
by a known potential $V_{L}$. This long-range interaction is made of meson
exchanges as described in section~\ref{II} and it contains of course also
the electromagnetic interaction.

A nice, instructive example is the {\it modified black disk},
where $V_{L} = 0$ and $P = -ipb$.
The $P$ matrix contains a negative imaginary part, implying
absorption of flux at the boundary. The boundary $b$ is a measure
for the annihilation radius. This modified black disk is
specified by only one parameter: the radius $b$.

When one looks how these boundary-condition
models have been used, then one sees that in these
extremely simple models every time only very few
parameters have been introduced. The conclusion is that such a
few-parameter model can fit possibly some data, but it will never
be able to fit all the available $\overline{N}\!N$ data, with
the same set of only a few parameters.

\subsection{Optical-potential model}   \label{III.3}
The optical-potential models have become quite an
industry in the $\overline{N}\!N$ community. The first such model
was from R. Bryan and R. Phillips in 1968~\cite{Bry68}. In the
optical-potential model the interaction between the antinucleon and the
nucleon is described by a complex potential from $r=0$ to infinity. 
For the basic potential one takes a meson-theoretic potential, obtained
from some known $N\!N$ potential by using the charge-conjugation
operation. Then, in order to get annihilation, to this potential is
added another complex potential,
\[
V(r) = (U-iW) f(r)\ .
\]
Here $U$ and $W$ are constants and $f(r)$ is some radial function. This radial
function can be the Woods-Saxon form, a Gaussian form, or even a square well.
Let us give you a DO-IT-YOURSELF-KIT called:
\begin{center}
{\it How to make your own optical potential?}
\end{center}
Instructions: \\
1. Look through the literature and decide which $N\!N$ potential
   your want to use. \\
2. Apply to this potential charge conjugation, so that you obtain
   the corresponding $\overline{N}\!N$ potential. \\
3. Pick your favored functional form for $f(r)$. 
   This will contain a range parameter $b$. 
   After you have made this choice, find some arguments,
   which sound like QCD, to justify this chosen form. \\
4. Pick one of the beautiful differential cross sections as measured by
   Eisenhandler \etal~\cite{Eis76}
   and adjust $U$, $W$, and $b$ such that a reasonable
   fit (at least at sight) is obtained for this particular cross section. \\
Your model is now a three-parameter model, which fits {\it some}
of the data (at least the Eisenhandler data at one energy)
reasonably well (at sight), but it cannot possibly fit 
{\it all} the $\overline{N}\!N$ data, because the
model does not have enough freedom.

After Bryan and Phillips many people have constructed
similar models (see \eg\ Refs.~\cite{Dov80,Koh86,Hip89,Car91}).
Also in Nijmegen we made such an optical-potential model, 
which we optimized by making a least-squares fit to our database,
which contained at that time $N_{d} = 3309$ data. Because we
actually performed a fit to all the $\overline{N}\!N$ data we 
think that we will have about the lowest $\chi^2$ of all the
available two- or three-parameter optical-potential models. For our
model $\chi_{\rm min}^{2} = 6\ 10^{9}$. This enormously large number is
NOT a printing error, but just an expression of the total failure of
such simple models.
It is, therefore, astonishing to see that regularly new
measurements from LEAR are compared to one or more of these
few-parameter optical-potential models (see \eg\ Ref.~\cite{Kun91}),
as if something can be learned from such a comparison!

There is only one group, the theory group of R. Vinh Mau
in Paris, that has seriously tried to fit     
all available $\overline{N}\!N$ data with an
optical-potential model. In 1982 they got a fit with
$\chi^{2}/N_{d} = 2.8$, where they compared with the then
available pre-LEAR data~\cite{Cot82}. In 1991 they published an
update~\cite{Pig91}, where they fitted now also the LEAR data.
For the real part of the potential they took the $G$-conjugated
Paris $N\!N$ potential~\cite{Lac80}.
Because the inner region of this $N\!N$ potential is treated
totally phenomenologically it is impossible to take that over
to $\overline{N}\!N$, so something has been done there and 
probably some extra parameters have been introduced.
The imaginary part of the potential they write as
\begin{eqnarray*}
W(r) & = & \left\{ \rule{0mm}{6mm} g_{c} (1+f_{c}T_{L})+g_{ss}(1+f_{ss}T_{L})\,
\mbox{\boldmath $\sigma$}_{1} \cdot \mbox{\boldmath $\sigma$}_{2} \right. \\
 && \left. +\, g_{T} S_{12} + g_{LS}\ {\bf L} \cdot {\bf S}\ \frac{1}{4m^{2}r}
   \frac{d}{dr} \right\} \frac{K_{0}(2mr)}{r} \ ,
\end{eqnarray*}
where $T_L$ is the lab kinetic energy and
the parameters are the $g$'s and the $f$'s. For each isospin a set
of 6 parameters is fitted, so that the imaginary part is described
by about 12 real parameters. In total the Paris $\overline{N}\!N$
potential uses at least 12, possibly about 22, parameters. 
The correct number used is not so important, what is
important, is that the number is much larger than 3. The Paris group
do fit then to 2714 data and get $\chi^{2}/N_{d} = 6.7$.
The quality of this fit is very hard to assess, because the Paris group
did not try to make their own selection of the data, but tried to fit
all the available data, many of which are contradictory. It would
be interesting to see their fit to the Nijmegen
$\bar{p}p$ database \cite{Tim93} (see section~\ref{VIII}), 
where all the contradictory sets have been removed.

An important lesson could have been learned already in 1982 from
this Paris work. An optical-potential model needs at least about
15 parameters to be able to give a reasonable fit to the 
$\overline{N}\!N$ data. This means that practically all
few-parameter optical-potential models published after 1982
should have been rejected by the journals.

\subsection{Coupled-channels model}  \label{III.4}
Another way to introduce inelasticity in our formalism is to
introduce explicitly couplings from the $\overline{N}\!N$
channels to annihilation channels. 
This was done in 1984 by P. Timmers \etal\ in the Nijmegen
coupled-channels model: CC84~\cite{Tim84}. Fitting to the then
available pre-LEAR data resulted in a quite satisfactory fit
with $\chi^{2}/N_{d} = 1.39$. Several people have
have later tried similar models~\cite{Liu90,Dal90}. An update of
the old model CC84 was made in 1991 in Nijmegen in the thesis of
R. Timmermans~\cite{Tim91c}.
This new coupled-channels model, which we would like
to call the Nijmegen model CC93, gives $\chi^{2}/N_{d} = 1.58$,
when fitted to $N_{d} = 3646$ data.
We will come back to this model somewhat later.

\section{Antiproton-proton partial-wave analysis}  \label{VII}
In Nijmegen we have for almost 15 years been busy with partial-wave
analyses of the $N\!N$ data. We have now developed rather sophisticated
and accurate methods to do these PWA's~\cite{Ber88,Ber90,Klo93}.
A few years ago we realized that it was possible to do a PWA
of all the available $\bar{p}p$ data in {\it exactly the same way} as
our $N\!N$ PWA. Before this realization we always thought that such a
PWA would be almost impossible in $\overline{N}\!N$. Luckily, it is
not impossible. We will try to give a short description of our
PWA~\cite{Tim93}.

In an energy-dependent partial-wave analysis one needs a model to describe
the energy dependence of the various partial-wave amplitudes. Our model
is a mixture of the boundary-condition model and the optical-potential
model. We choose the boundary at $b=1.3$ fm. This value is
determined by the width of the diffraction peak and cannot be
chosen differently, without deteriorating the fit to the data.
The long-range potential $V_{L}$ for $r>b$ is
\[
V_L = V_{\overline{N}\!N} + V_{C} + V_{M\!M}\ .
\]
Here $V_{C}$ is the relativistic Coulomb potential, $V_{M\!M}$ the
magnetic-moment interaction, and $V_{\overline{N}\!N}$ is
the charge-conjugated Nijmegen $N\!N$ potential, Nijm78~\cite{Nag78}.
We solve the relativistic Schr\"odinger equation~\cite{Swa78} for each
energy and for each partial wave, subject to the boundary condition
\[
P = b \left( \frac{d\psi}{dr}/\psi \right)_{r=b} \ ,
\]
at $r=b$.
This boundary condition may be energy dependent. To get the value of $P$
as a function of the energy we use for the spin-uncoupled waves
(like $^{1}S_{0}$, $^{1}P_{1}$, $^{1}D_{2}$, $\ldots$ and $^{3}P_{0}$, 
$^{3}P_{1}$, $^{3}D_{2}$, $\ldots$) the optical-potential picture.
We take a square-well optical potential for $r \leq b$. This
short-ranged potential $V_{S}$ we write as
\[
V_{S} = U_{S} - iW_{S}\ .
\]
In this way we get in each partial wave and for each isospin the 
parameters $U_{S}$ and $W_{S}$. Using these potentials we can calculate
easily the boundary condition $P$ and the scattering amplitudes.
For example in all singlet waves $^{1}S_{0}$, $^{1}P_{1}$, $^{1}D_{2}$, 
$\ldots$, we get $U=0$ and $W \approx 100$ MeV. 
For the triplet waves we take $W$
independent of the isospin. The parameters for the
$^{3}P_{0}$ wave are \eg\ $W=159 \pm 9$ MeV and independent
of the isospin, and
$U (I=0) = -132 \pm 9$ MeV and $U (I=1) = 178 \pm 19$ MeV.
To describe all relevant partial waves we need in our $\overline{N}\!N$
PWA 30 parameters. In our fit to the data we use all
available data in the momentum interval
119 MeV/c $< p_{L} <$ 923 MeV/c. The lowest
momentum is determined by the fact that for lower momenta no
data are available. The highest momentum is determined by several
considerations. In $N\!N$ we use all data up to $T_{L} = 350$ MeV,
which corresponds to $p_{L} = 810$ MeV/c.
Because we wanted to include all the elastic backward 
cross sections of Alston-Garnjost \etal~\cite{Als79},
we need to go to $p_{L} = 923$ MeV/c which corresponds to
$T_{L} = 454$ MeV. At this energy the potential description in $N\!N$ is
still valid, and therefore we feel that also here our description
must work at least up to this momentum.

Our final dataset contains $N_{d}=3646$ experimental data. In our 
analyses we need to determine $N_{n}=113$ normalizations and $N_{p}=30$
parameters. This leads to the number $N_{df}$ of degrees of freedom
$N_{df} = N_{d} - N_{n} - N_{p} = 3503$.
When the dataset is a perfect statistical ensemble and when the model
to describe the data is totally correct, then one expects for
$\chi^2_{\rm min}$:
\[
\langle \chi^{2}_{\rm min} \rangle = N_{df} \pm \sqrt{2N_{df}}\ .
\]
Thus expected is $\langle \chi^{2}_{\rm min}(\bar{p}p) 
\rangle/N_{df} = 1.000 \pm .024$. \\
In our PWA we obtain $\chi^{2}_{\rm min}(\bar{p}p)/N_{df} = 1.085$.
We see that we are about 3.5 standard deviations
away from the expectation value. To get a feeling
for these numbers let us compare with the $pp$ data and the
$pp$ PWA. The number of data is now $N_{d} = 1787$ and we expect
\[
\langle \chi^{2}_{\rm min}(pp) \rangle /N_{df} = 1.000 \pm 0.035\ .
\]
In the latest Nijmegen analysis, NijmPWA93~\cite{Klo93}, we get
\[
\chi^{2}_{\rm min}(pp)/N_{df} = 1.108,
\]
which is 3 standard deviations too high. We see
that our $\bar{p}p$ analysis compares favorable 
with a similar analysis for the $pp$ data.
This means therefore that we have a statistically rather good
solution and also that this solution will be essentially correct.

\section{The Nijmegen ${\bf \overline{N}\! N}$ database} 
   \label{VIII}
An essential ingredient in our successfully completed PWA,
as well as an important product of this PWA, is the Nijmegen
$\overline{N}\!N$ database~\cite{Tim93}.
As pointed out before, we use all data with $p_{L} < 925$ MeV/c
or $T_{L} < 454$ MeV. This means that our momentum range is
similar to the momentum range used in the Nijmegen $N\!N$ PWA's.
We will compare regularly with the $N\!N$ case to show that the
same methods, which work well in $N\!N$, work also well in $\overline{N}\!N$
and that the results are also similar.

The number of data $N_{d}$ in  the various final datasets are
$N_{d}(\bar{p}p) = 3543$, $N_{d}(pp) = 1787$, and $N_{d}(np) = 2514$. 
In the processes to come to these final datasets we had to reject data.
We do not want to go into details~\cite{Ber88}
about what are the various criteria
to remove data from the dataset. We would like to point out, however,
that in $pp$ scattering there is a long history about which datasets
are reliable, and which not. We did not invent the method of discarding
incorrect data, we just followed common practice and used common sense.
In the $\bar{p}p$ case we needed to reject 744 data, which is 17\% of
our final dataset. 
In the $pp$ case we discarded 292 data or 14\% of the final dataset,
and in the $np$ case we rejected 932 data, which amounts to 27\%
of the final dataset. 
It is clear that the $\bar{p}p$ case does not seem to be out of bounds.
Of course, it is unfortunate that so many data have to be rejected,
because these data represented many man-years of work and a lot of
money and effort. However, when one wants to treat the data in a
statistically correct manner, then often one cannot handle all
datasets, but one must reject certain datasets. 
This does not mean that all these rejected datasets
are ``bad'' data, it only means, that if we want to apply statistical
methods, then, unfortunately, certain datasets cannot be used.     

\begin{table}
\centering
\begin{tabular}{lcc|cc}
 & \multicolumn{2}{c|}{elastic} & \multicolumn{2}{c}{charge exchange} \\
 & LEAR & rest & LEAR & rest \\ \hline
$\sigma_{T},\sigma_{A}$ & 124 & -    & -  & 63 \\
$\sigma(\theta)$        & 281 & 2507 & 91 & 154 \\
$A(\theta)$             & 200 & 29   & 89 & - \\
$D$                     & 5   & -    & 9  & - \\ \hline
\multicolumn{1}{l}{total}  & 610 & 2536 & 189 & 217 
\end{tabular}
\caption{Number of elastic and charge-exchange data
         divided over various categories.}  \label{tab.I}
\end{table}

In Table~\ref{tab.I} we give the number of data points
divided into elastic versus charge exchange,
LEAR versus the rest, and total cross sections
$\sigma_{T}$, annihilation cross sections $\sigma_{A}$, 
differential cross sections $\sigma(\theta)$, analyzing-power
data $A(\theta)$, and depolarization data $D(\theta)$. This table
give some interesting information. \\
The most striking fact is that: \\
{\it Of the final dataset only $22$\% of the data comes from LEAR.} \\
This is after 10 years operation of LEAR. Remember the promises
(or were it boasts)
from CERN, made before LEAR was built. They were something like:
``Only one day running of LEAR will produce
more scattering data then all other methods together.''
Unfortunately, this promise of CERN did not work out.
Also it is clear that LEAR has not given much valuable information
about $\sigma(\theta)$ for the elastic reaction.
A lot of the elastic $\sigma(\theta)$ data from
LEAR unfortunately needed to be rejected~\cite{Tim93}! 
This does not mean that LEAR did not produce beautiful data. Some of
the charge-exchange data and the strangeness-exchange data are really
of high quality. 

Another striking fact is the virtual absence of spin-transfer and
spin-correlation data. For the elastic reaction below 925 MeV/c
there are only 5 depolarizations measured with enormous
errors~\cite{Kun91}. Very recently, some depolarization
data of good quality have become available for the
charge-exchange reaction~\cite{Bir93}.

A valid question is therefore: \\
{\it Can one do a PWA of the $\bar{p}p$ data, when there are
essentially no ``spin data''?} \\
The answer is yes! The proof that it can be done lies in the fact that we
actually produced a $\bar{p}p$ PWA with a very good $\chi^{2}/N_{d}$.
We have also checked this at length in our $pp$ PWA's.
We convinced ourselves that a $pp$ PWA using only $\sigma(\theta)$ and
$A(\theta)$ data gives a pretty good solution. Of course,
adding spin-transfer and spin-correlation data
was helpful and tightened the error bands. However, most
spin-transfer and spin-correlation data in the 
$pp$ dataset actually did not give any additional information.

\section{Fits to the data}         \label{IX}
\begin{figure}
\vspace*{8cm}
 \includegraphics{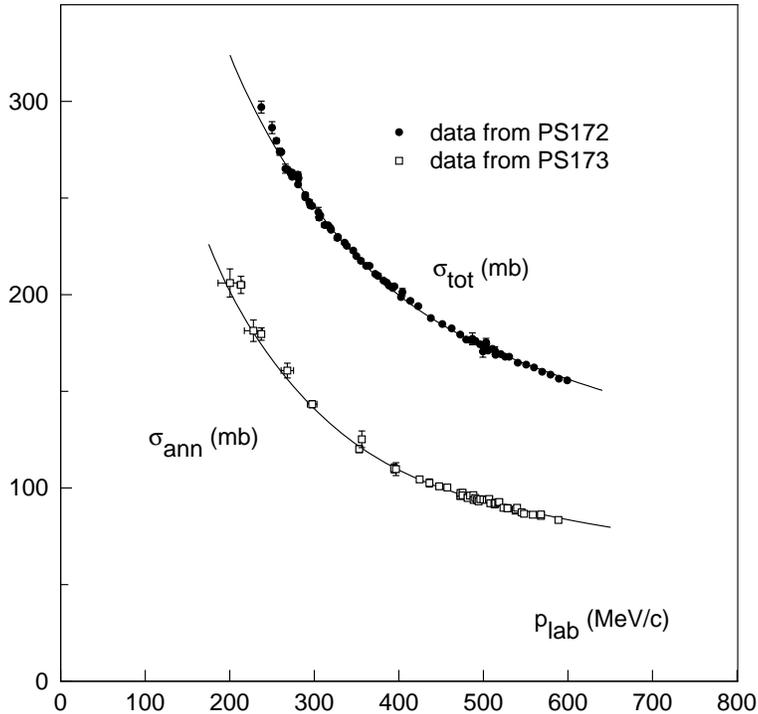}
\caption{Total cross sections (PS172) and annihilation cross
         sections (PS173) with the curves from the Nijmegen PWA.}
\label{fig.1}
\end{figure}

\begin{figure}
\vspace*{8cm}
 \includegraphics{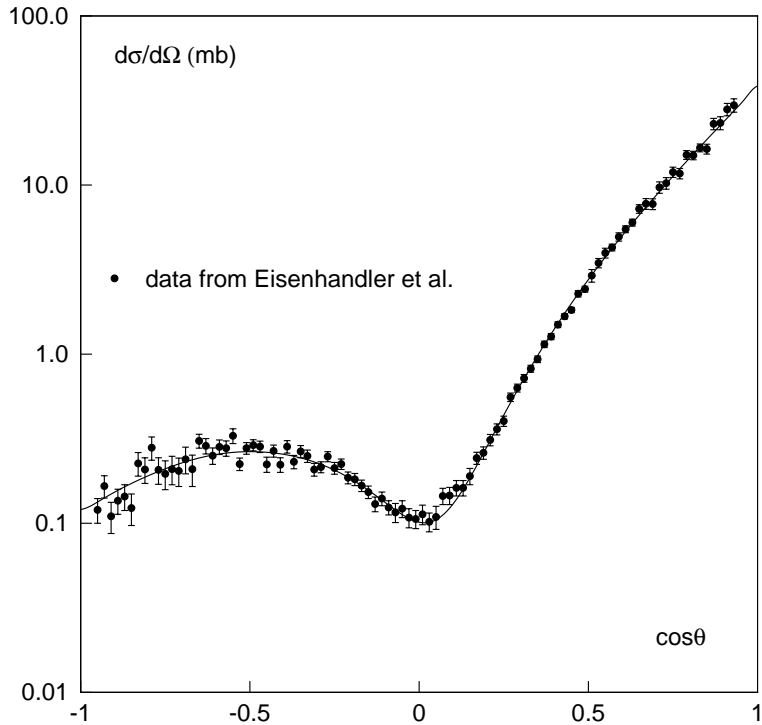}
\caption{Elastic differential cross section at 790 MeV/c from Eisenhandler
         {\it et al}., with the curve from the Nijmegen PWA.}
\label{fig.2}
\end{figure}

It is of course impossible to show here how well the various
experimental data are fitted. From our final $\chi^{2}/N_{d} = 1.085$
one can draw the conclusion, that almost every dataset will have a
contribution to $\chi^{2}$ which is roughly equal to the
number of data points as is required by statistics.
Let us look at some of the experimental data.
In Fig.~\ref{fig.1} we present total cross sections from PS172~\cite{Clo84}.
The fit gives for these 75 data points $\chi^{2} = 88.4$.
In the same Fig.~\ref{fig.1} one can also find 52 annihilation
cross sections from PS173~\cite{Bru87}.
These points contributed $\chi^{2} = 65.3$
to the total $\chi^{2}$.
In Fig.~\ref{fig.2} we plot the elastic
differential cross section $\sigma(\theta)$ at 
$p_L = 790$ MeV/c as measured in 1976 by Eisenhandler \etal~\cite{Eis76}
The 95 data points contribute $\chi^{2} = 101.5$. The vertical scale
is logarithmic. The nice fit reflects the high quality of these
pre-LEAR data.

\begin{figure}
\vspace*{8cm}
 \includegraphics{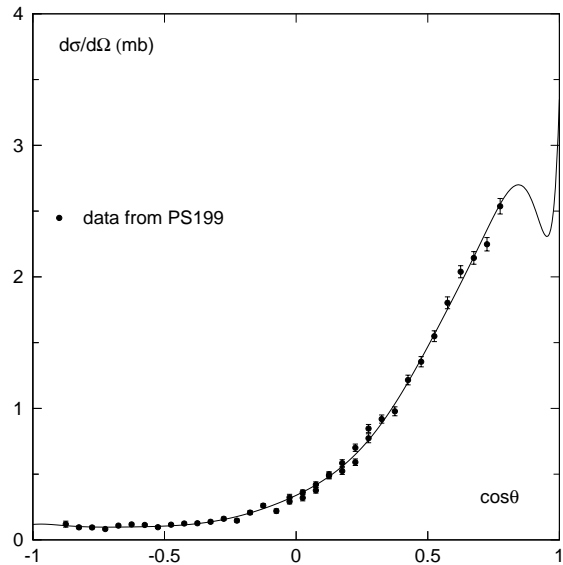}
\caption{Charge-exchange differential cross section at 693 MeV/c
         from PS199, with the curve from the Nijmegen PWA.}
\label{fig.3}
\end{figure}

The differential cross section of the charge-exchange reaction
$\bar{p}p \rightarrow \bar{n}n$ at \linebreak[4]
$p_{L}=693$ MeV/c as measured
by PS199 is given in Fig.~\ref{fig.3}. The 33 data contribute \linebreak[4]
$\chi^{2} = 39.3$. This dataset can be considered important,
because it is very constraining. One needs all partial waves up
to $L=10$ to get a satisfactory fit to these data.

\section{Coupled-channels potential model} \label{X}
Having finished our discussion of the Nijmegen $\bar{p}p$
partial-wave analysis we can look at the $\overline{N}\!N$ potentials.
We decided to update the old coupled-channels model Nijmegen CC84 of
Timmers \etal~\cite{Tim84}. Because of our experience with the various
datasets this was not very difficult, just very computer-time consuming.
The result was the new Nijmegen CC93 model~\cite{Tim91c}.
In this model we treat the $\overline{N}\!N$ coupled channels on the
particle basis. We therefore have a $\bar{p}p$ channel as well as a
$\bar{n}n$ channel. This allows us to introduce the charge-independence
breaking effects of the Coulomb interaction in the
$\bar{p}p$ channel and of the mass differences between the
proton and neutron as well as between the exchanged $\pi^{0}$ and
$\pi^{\pm}$.

These $\overline{N}\!N$ channels are coupled to annihilation channels.
We assume here that annihilation can happen only into two fictitious mesons;
into one pair of mesons with total mass 1700 MeV/c$^{2}$, and into
another pair with total mass 700 MeV/c$^{2}$. Moreover, we assume
that these annihilation channels appear in both isospins $I=0$
as well as $I=1$.
We end up with 6 coupled channels for each of the $\bar{p}p$ channels:
$^{1}S_{0}$, $^{1}P_{1}$, $^{1}D_{2}$, $^{1}F_{3}$, etc.\
and $^{3}P_{0}$, $^{3}P_{1}$, $^{3}D_{2}$, $^{3}F_{3}$, etc.
Due to the tensor force we end up with 12 coupled channels
for each of the $\bar{p}p$ coupled channels:
$^3S_1+\,^3D_1$, $^3P_2+\,^3F_2$, 
$^3D_3+\,^3G_3$, etc.

We use the relativistic Schr\"odinger equation in coordinate space.
The interaction is then described by either a $6\times 6$ or a
$12\times12$ potential matrix
\[
V = \left( \begin{array}{cc}
   V_{\overline{N}\!N} & V_{\!A} \\[2mm]
   \widetilde{V}_{\!A} & 0
    \end{array} \right) \ .
\]
The $2\times 2$ (or $4\times 4$) submatrix $V_{\overline{N}\!N}$
we write as
\[
V_{\overline{N}\!N} = V_{C} + V_{M\!M} + V_{O\!B\!E}\ ,
\]
where for $V_{C}$ we use the relativistic Coulomb potential, $V_{M\!M}$
describes the magnetic-moment interaction, and for $V_{O\!B\!E}$ we use
the charge-conjugated Nijmegen $N\!N$ potential Nijm78~\cite{Nag78}.
We have assumed that we may neglect the diagonal interaction in the
annihilation channels. The annihilation potential $V_{\!A}$
connects the $\overline{N}\!N$ channels to the two-meson annihilation 
channels. It is either a $2\times 4$ matrix or a $4\times 8$ matrix.
This potential we write as
\[
V_{\!A}(r) = \left( V_{C} + V_{SS} \mbox{\boldmath $\sigma$}_{1} \cdot
 \mbox{\boldmath $\sigma$}_{2} + V_{T} S_{12} m_{a}r + V_{SO}
  {\bf L}\cdot {\bf S} \frac{1}{m_{a}^{2}r} \frac{d}{dr} \right)
  \frac{1}{1+e^{m_ar}}\ .
\]
The factor $m_{a}r$ is introduced in the tensor force to make
this potential identically zero at the origin. Here $m_{a}$ is the
mass of the meson (either 850 MeV/c$^{2}$ or
350 MeV/c$^{2}$).
This annihilation potential depends on the spin structure of the initial
state. For each isospin and for each meson channel five parameters
are introduced: $V_{C}$, $V_{SS}$, $V_{T}$, $V_{SO}$, and $m_a$.
This gives a model with in total $4\times 5=20$ parameters. These
parameters can then be fitted to the $\overline{N}\!N$ data.
Doing this we obtained $\chi^{2}/N_{d}=3.5$.
It is clear, of course, that although the old Nijmegen soft-core
potential Nijm78 is a pretty good $N\!N$ potential, it is
definitely not the ultimate potential. We decided therefore to
introduce now as extra parameters the coupling constants of the 
$\rho$, $\omega$, $\varepsilon(760)$, 
pomeron, and $a_{0}(980)$. Adding these parameters
allowed us quite a drop in $\chi^{2}$. Now we reached
\[
\chi^{2}/N_{d} = 1.58 \ ,
\]
with a total of 26 parameters.

\section{The reaction $\bar{p}p \rightarrow \bar{\Lambda}\Lambda$} \label{XI}
It is perhaps not superfluous to point out here, that we also made a
PWA of the strangeness-exchange reaction $\bar{p}p \rightarrow \bar{\Lambda}
\Lambda$~\cite{Tim91a,Tim92}.
Fitting the $N_{d}=142$ data, we get $\chi^{2}_{\rm min}/N_{d}=1.027$.

The first theoretical treatments of this reaction were by F. Tabakin
and R.A. Eisenstein~\cite{Tab85} and independently by P. Timmers
in his thesis~\cite{Tim85}.
Many other treatments of this reaction can be found (see
{\it e.g.} Refs.~\cite{Nis85,Koh86a,Fur87,Kro87,Alb88,LaF88,Hai93}).
In the meson-exchange
models it is clear that next to $K(494)$ exchange, there is also
the exchange of the vector meson $K^{*}(892)$. In Nijmegen we have
been able to determine the $\Lambda N\!K$ coupling constant at the
pole~\cite{Tim91a}. We found \linebreak[4]
$f^{2}_{\Lambda N\!K} = 0.071 \pm 0.007$. This value is in agreement with the
value $f^{2}_{\Lambda N\!K}=0.0734$ \linebreak[4]
used in the recent soft-core Nijmegen
hyperon-nucleon potential~\cite{Mae89}. When we determine also the mass of
the exchanged pseudoscalar meson we find \linebreak[4]
  $m(K) = 480 \pm 60$ MeV in
good agreement with the experimental value \linebreak[4]
 $m(K)=493.646(9)$ MeV. 
This shows that we are actually looking at the one-kaon-exchange mechanism
in the reaction $\bar{p}p \rightarrow \bar{\Lambda}\Lambda$. 

When the data for the reactions $\bar{p}p \rightarrow \bar{\Lambda}\Sigma$
and $\bar{\Sigma}\Lambda$ are available, then also the $\Sigma N\!K$ coupling
constant can be determined. When this can be done with sufficient
accuracy, then information about the SU(3) ratio $\alpha = F/(F+D)$ can be
obtained, and SU(3) for these coupling constants can then actually be
studied.

\section{PWA as a TOOL}  \label{concl}
We have presented here some of the results of the first, energy-dependent
partial-wave analysis of the elastic and charge-exchange $\bar{p}p$
scattering data~\cite{Tim93}. We also discussed the Nijmegen 
$\overline{N}\!N$ dataset, where we removed the contradictory or otherwise
not so good data from the world $\bar{p}p$ dataset. 

The main reason that we have been able to perform  a PWA of the
$\bar{p}p$ scattering data is that practically all partial-wave amplitudes
are dominated by the potential outside $r=1.3$ fm. This long range 
potential consists of the electromagnetic potential, the OPE potential,
and the exchange potentials of the mesons like $\rho,\ \omega,\
\varepsilon$, etc. This long-range potential is therefore well known.
In our PWA of the $N\!N$ scattering data~\cite{Klo93} it was noticed by
us that the long-range potential in the $N\!N$ case dominated the
$N\!N$ partial-wave scattering amplitudes. In the $\bar{p}p$ case the
long-range potential is much stronger (see section~\ref{II}), and
the dominance in the $\bar{p}p$ case is therefore more marked.
One could formulate this the following way. The important $\bar{p}p$
partial-wave amplitudes are ``$\pi,\ \rho,\ \varepsilon$, and $\omega$
dominated.'' This gives the most important energy dependence of
these amplitudes. The slower energy dependence due to the short-range
interaction can easily be parametrized. 

A second reason for the successful PWA is the availability and 
easy access to computers, because the methods used are very computer
intensive. 

We want to stress the fact that our multienergy PWA can now be used
as a {\bf tool}. This tool allows us first of all to judge the 
quality of a particular dataset. This enabled to us to set up the Nijmegen
$\overline{N}\!N$ database. Secondly, it can be used in the study
of the $\overline{N}\!N$ interaction.

To demonstrate these things let us look at the Meeting Report
of the Archamps meeting from October 1991~\cite{Bra93}. Beforehand
the participants were asked to discuss at the meeting such questions as: \\
{\it What is the evidence for one-pion exchange in the 
 $\overline{N}\!N$ interaction?} \\
In Nijmegen we determined~\cite{Tim91b,Tim93}, using the PWA
as a tool, the $N\!N\pi$ coupling constant for charged pions from
the data of the charge-exchange reaction $\bar{p}p \rightarrow \bar{n}n$.
We found~\cite{Tim93}
\[
f_{c}^{2} = 0.0732 \pm 0.0011\ .
\]
This is only 64 standard deviations away from zero!!
Using analogous techniques we could also
determine this coupling constant for charged pions in our analyses
of the $np$ scattering data. 
We found there~\cite{Sto93}
\[
f_{c}^{2} = 0.0748 \pm 0.0003\ .
\]
The same coupling constant can also be seen in analyses of the
$\pi^{\pm}p$ scattering data. There the VPI\&SU group
finds $f_{c}^{2} = 0.0735\pm 0.0015$~\cite{Arn90}.
In $pp$ scattering we have
determined the $pp\pi^{0}$ coupling constant.
Our latest determination gives~\cite{Sto93}
\[
f_{p}^{2} = 0.0745 \pm 0.0006\ .
\]
The nice agreement between these different values shows \\
(1) the charge independence for these coupling constants and its shows that\\
(2) the presence of OPE in the $\overline{N}\!N$ interaction 
is a 64 s.d.\ effect. \\
{\it What more evidence does one wants?} \\
We also played around with the pion masses. 
In $N\!N$ scattering we were able to determine
the masses of the $\pi^{0}$ and $\pi^{\pm}$. We found there
$m_{\pi^0}=135.6(1.3)$ and $m_{\pi^\pm}=139.4(1.0)$ MeV/c$^2$,
to be compared to the particle-data values \\
$m_{\pi^0}=134.9739$ and $m_{\pi^\pm}=139.56755$ MeV/c$^2$. We did
not try to determine these masses again in $\overline{N}\!N$ scattering.
However, we think we could have. We checked that changing the correct
pion masses to an averaged $\pi$-mass raised our
$\chi^{2}_{\rm min}(\bar{p}p)$ with 9.

Another question posed before that meeting was
{\it ``What is the evidence for the $G$-parity rule?''}
In our determination of the $N\!N\pi$ coupling constant in the
charge-exchange reaction this G-parity rule was of course implicitly 
assumed. Our determination of $f_{c}^{2}$ and its agreement with
the expected value can therefore be seen as a proof of this rule
for pion exchange.

When one looks through the literature one finds several, what we think,
artificially created problems. Why is this done? Only to get beamtime?
One of such problems is the statement: ``The OBE model does not work.''
We would like to point out that the OBE model works
excellently~\cite{Tim91c}. Other
examples can be found in the already mentioned Archamps Meeting
Report~\cite{Bra93}. The authors of this report claim
that the charge-exchange differential cross sections at low
energy pose a {\it challenge for every model}. Let us look at those
data. Contrary to what is stated in the Meeting Report these data
are a part of our dataset, so we have sufficient knowledge to discuss
them. The discussion concerns data of PS173~\cite{Bru86b}.
At four momenta the differential
cross section for $\bar{p}p \rightarrow \bar{n}n$ 
was measured. The results of our PWA for these measurements are: \\
At $p_{L} = 183$ MeV/c there are 13 $d\sigma_{ce}/d\Omega$ data. 4 of
these data are rejected because each of them contributes more than 9
to our $\chi^{2}$. This is the three-standard-deviation rule.
The remaining 9 data contribute $\chi^{2}=8.3$. \\
At $p_{L}=287$ MeV/c there are 14 $d\sigma_{ce}/d\Omega$ data, where 1
of these data points is discarded because it contributes more than 9
to our $\chi^{2}$. The remaining 13 data contribute
$\chi^{2}=24.0$. \\
At $p_{L}=505$ MeV/c there are 14 $d\sigma_{ce}/d\Omega$ data. One
of them is discarded because of its too large $\chi^{2}$ contribution.
The remaining 13 data contribute $\chi^{2}=30.1$. \\
At $p_{L}=590$ MeV/c there are 15 $d\sigma_{ce}/d\Omega$ data, where 2  
of them are discarded.
The remaining 13 data points contribute $\chi^{2}=32.8$.

What can we conclude? At the lowest momentum we rejected 30\% of the
data and the remaining dataset is then OK. However, at the other three
momenta we find rather large contributions to $\chi^{2}$. A dataset
of 13 data is, according to the three-standard-deviation rule, not
allowed to contribute more than $\chi^{2}_{\rm max}=31.7$
to the $\chi^{2}_{\rm min}$ of our database. 
This means that we really should reject the data at $p_{L} = 590$
MeV/c. When we combine the 4 datasets to one dataset with 47 data points,
we see that these data points contribute $\chi^{2}=95.2$ to
$\chi^{2}_{\rm min}$ of our database. The rule says that a set
of 47 data may not have a $\chi^{2}$-contribution larger than
78.5. This means that this whole dataset should be rejected.
The only reason, that these dubious data are still contained in the
Nijmegen $\overline{N}\!N$ database and not discarded, is that
there are no other charge-exchange data at such low momenta.
Our philosophy here was that these
imperfect data are perhaps better than no data at all. The authors of
the Archamps Meeting Report~\cite{Bra93}, two experimentalists and a
phenomenologist, are obviously incorrect. Our PWA shows clearly that these
data cannot pose ``a challenge for every model,'' because these data
should really be discarded!

Another {\it challenge for models} seems to be that
{\it ``the strangeness-exchange reaction}
$\bar{p}p \rightarrow \bar{\Lambda}\Lambda$ {\it takes place in
almost pure triplet states.''} Let us look for a moment
in more detail at the beautiful data of PS185~\cite{Bar87}.
These data have been studied by many people. In Nijmegen we
performed also a PWA of these data~\cite{Tim91a,Tim92}.
It is very clear from our PWA that in this reaction the tensor force
plays a dominant role. The tensor force acts only in triplet waves.
These triplet waves make up the bulk of the cross section. This result
has been confirmed by several groups and clearly this is not a challenge,
but only a case of strong tensor forces. In section~\ref{II} we already
explained the reason for such strong tensor forces.

\begin{figure}
\vspace*{8cm}
 \includegraphics{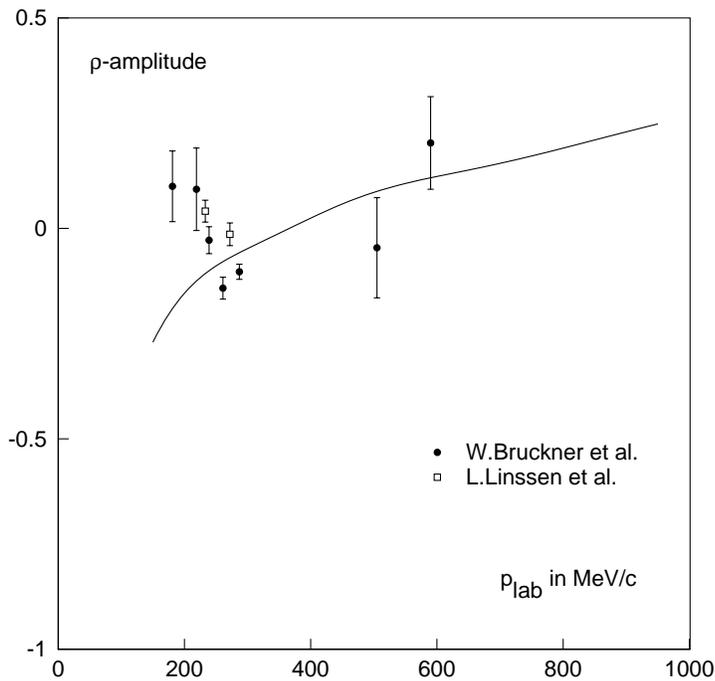}
\caption{The $\rho$ amplitude. Data are from PS172 and PS173. The
         curve is the prediction from the Nijmegen PWA.}
\label{fig.4}
\end{figure}

A big deal is often made of the $\rho$ parameter, the
real-to-imaginary ratio of the forward scattering amplitude.
The extraction of
this parameter from the available experimental data is based on a rather
shaky theory and on not much better data, polluted by Moli\`ere
scattering and, in our opinion, the underestimation of systematic errors.
When we look for example at the seven $\rho$ determinations by
PS173~\cite{Bru85} then we note that this
group has published only at four of these energies the
corresponding $d\sigma/d\Omega$ data~\cite{Bru86a}!
In our PWA we discard these data at
three of the energies. We feel therefore strongly, that the  
$\rho$ determinations by PS173 should clearly be discarded and very
probably the errors on the determinations by PS172~\cite{Lin87}
should be enlarged considerably.
This leads to the simple picture as shown in Fig.~\ref{fig.4}. 

Another curious trend is the direct comparison between predictions
of meson-exchange models and of simple quark-gluon models for
the strangeness-exchange reaction. There are even serious 
proposals~\cite{Hai92,Alb93} for experiments to distinguish between 
these models: it is proposed to measure the spin transfer in $\bar{p}p
\rightarrow \bar{\Lambda}\Lambda$.

Let us make it clear from the outset, that we believe that {\bf all}
data must eventually be explained in terms of quark-gluon exchanges,
because this is the underlying theory. However, for the analogous
$N\!N$ interaction one has unfortunately not yet succeeded to give
a proper explanation of the meson-exchange mechanism in terms of quarks
and gluons exchanges. The theory is not so advanced yet. In the 
$\overline{N}\!N$ reactions we are of course in exactly the same
situation. Using our PWA as a tool we determined the $\Lambda N\!K$ coupling
constant {\bf and} the mass of the exchanged kaon. This way we established
beyond {\bf any} doubt that the one-kaon-exchange potential is present
in the transition potential and that again the tensor potential dominates.
It is absolutely not necessary to measure the spin transfer to
distinguish between the $K(494)$- and $K^{*}(892)$-exchange picture and
a simple quark-gluon-exchange picture. This distinction has already
been made using our PWA as a tool and using just the differential
cross sections and polarizations.

It has become a fad to promote the measurements of spin-transfer and
spin-correlation data, as if these data will solve all our troubles. For
example in the already often mentioned Archamps Meeting Report~\cite{Bra93}
one can read that the longitudinal spin transfer is obviously a favorite of
one of its authors. Somewhere else in the same report one finds the
question: ``Which new spin measurement would be crucial to confirm or
rule out present models?'' The answer to this last question is of
course: None! When one measures differential cross sections, polarizations,
spin transfers, or spin correlations, carefully enough, none of the present
models will fit these new data, but adjustments will be made in the models
in such a way that they do fit the data again. Physics is hard work
from experimentalists as well as from theorists. One needs many
and varied data and one single experiment has only a marginal
influence.

\vspace{\baselineskip}

\noindent {\bf Acknowledgments} \\
Part of this work was included in the research program of the Stichting voor
Fundamenteel Onderzoek der Materie (FOM) with financial support from the
Nederlandse Organisatie voor Wetenschappelijk Onderzoek (NWO).


\begin{thebibliography}{99}
\bibitem{Kun88}     R.A. Kunne {\it et al}.,
                    Phys. Lett. B {\bf 206}, 557 (1988);
                    Nucl. Phys. {\bf B323}, 1 (1989).
\bibitem{Ber89}     R. Bertini {\it et al}.,
                    Phys. Lett. B {\bf 228}, 531 (1989);
                    F. Perrot {\it et al}.,
                    {\it ibid}. {\bf 261}, 188 (1991).
\bibitem{Kun91}     R.A. Kunne {\it et al}.,
                    Phys. Lett. B {\bf 261}, 191 (1991).
\bibitem{Tim91b}    R. Timmermans, Th.A. Rijken, and J.J. de Swart,
                    Phys. Rev. Lett. {\bf 67}, 1074 (1991).
\bibitem{Bir90}     R. Birsa {\it et al}.,
                    Phys. Lett. B {\bf 246}, 267 (1990);
                    {\it ibid}. {\bf 273}, 533 (1991).
\bibitem{Bir93}     R. Birsa {\it et al}.,
                    Phys. Lett. B {\bf 302}, 517 (1993).
\bibitem{Bar87}     P.D. Barnes {\it et al}.,
                    Phys. Lett. B {\bf 189}, 249 (1987);
                    {\it ibid}. {\bf 199}, 147 (1987);
                    {\bf 229}, 432 (1989);
                    Nucl. Phys. {\bf A526}, 575 (1991).
\bibitem{Bar90}     P.D. Barnes {\it et al}.,
                    Phys. Lett. B {\bf 246}, 273 (1990).
\bibitem{Tim91a}    R. Timmermans, Th.A. Rijken, and J.J. de Swart,
                    Phys. Lett. B {\bf 257}, 227 (1991).
\bibitem{Pai52}     A. Pais and R. Jost, Phys. Rev. {\bf 87}, 871 (1952);
                    C. Goebel, Phys. Rev. {\bf 103}, 258 (1956);
                    T.D. Lee and C.N. Yang, 
                    Il Nuovo Cimento {\bf 3}, 749 (1956).
\bibitem{Spe67}     M.S. Spergel,
                    Il Nuovo Cimento {\bf 47A}, 410 (1967).
\bibitem{Dal77}     O.D. Dal'karov and F. Myhrer,
                    Il Nuovo Cimento {\bf 40A}, 152 (1977).
\bibitem{Del78}     A. Delville, P. Jasselette, and J. Vandermeulen,
                    Am. J. Phys. {\bf 46}, 907 (1978).
\bibitem{Bry68}     R.A. Bryan and R.J.N. Phillips,
                    Nucl. Phys. {\bf B5}, 201 (1968); {\it ibid}.
                    {\bf B7}, 481(E) (1968).
\bibitem{Eis76}     E. Eisenhandler {\it et al}.,
                    Nucl. Phys. {\bf B113}, 1 (1976).
\bibitem{Dov80}     C.B. Dover and J.-M. Richard,
                    Phys. Rev. C {\bf 21}, 1466 (1980);
                    {\it ibid}. {\bf 25}, 1952 (1982).
\bibitem{Koh86}     M. Kohno and W. Weise,
                    Nucl. Phys. {\bf A454}, 429 (1986).
\bibitem{Hip89}     T. Hippchen, K. Holinde, and W. Plessas,
                    Phys. Rev. C {\bf 39}, 761 (1989).
\bibitem{Car91}     J. Carbonell, O.D. Dal'karov, K.V. Protasov,
                    and I.S. Shapiro,
                    Nucl. Phys. {\bf A535}, 651 (1991).
\bibitem{Cot82}     J. C\^ot\'e, M. Lacombe, B. Loiseau,
                    B. Moussallam, and R. Vinh Mau,
                    Phys. Rev. Lett. {\bf 48}, 1319 (1982).
\bibitem{Pig91}     M. Pignone, M. Lacombe, B. Loiseau, and
                    R. Vinh Mau,
                    Phys. Rev. Lett. {\bf 67}, 2423 (1991).
\bibitem{Lac80}     M. Lacombe, B. Loiseau, J.-M. Richard, R. Vinh Mau,
                    J. C\^ot\'e, P. Pir\`es, and R. de Tourreil,
                    Phys. Rev. C {\bf 21}, 861 (1980).
\bibitem{Tim84}     P.H. Timmers, W.A. van der Sanden, and
                    J.J. de Swart,
                    Phys. Rev. D {\bf 29}, 1928 (1984).
\bibitem{Liu90}     G.Q. Liu and F. Tabakin,
                    Phys. Rev. C {\bf 41}, 665 (1990).
\bibitem{Dal90}     O.D. Dal'karov, J. Carbonell, and K.V. Protasov,
                    Sov. J. Nucl. Phys. {\bf 52}, 1052 (1990),
                    (Yad. Fiz. {\bf 52}, 1670 (1990)).
\bibitem{Tim91c}    R. Timmermans, Ph.D. thesis, University
                    of Nijmegen (1991); R. Timmermans, Th. A. Rijken,
                    and J.J. de Swart, in preparation.
\bibitem{Ber88}     J.R. Bergervoet, P.C. van Campen,
                    W.A. van der Sanden,  and J.J. de Swart,
                    Phys. Rev. C {\bf 38}, 15 (1988).
\bibitem{Ber90}     J.R. Bergervoet, P.C. van Campen, R.A.M. Klomp,
                    J.-L. de Kok, T.A. Rijken, V.G.J. Stoks, and
                    J.J. de Swart,
                    Phys. Rev. C {\bf 41}, 1435 (1990).
\bibitem{Klo93}     V.G.J. Stoks, R.A.M. Klomp,
                    M.C.M. Rentmeester, and J.J. de Swart,
                    Phys. Rev. C {\bf 48}, 792 (1993).
\bibitem{Tim93}     R. Timmermans, Th.A. Rijken, and
                    J.J. de Swart, submitted for publication.
\bibitem{Nag78}     M.M. Nagels, T.A. Rijken, and J.J. de Swart,
                    Phys. Rev. D {\bf 17}, 768 (1978).
\bibitem{Swa78}     J.J. de Swart and M.M. Nagels,
                    Fortschr. Phys. {\bf 26}, 215 (1978).
\bibitem{Als79}     M. Alston-Garnjost {\it et al}.,
                    Phys. Rev. Lett. {\bf 43}, 1901 (1979).
\bibitem{Clo84}     A.S. Clough {\it et al}.,
                    Phys. Lett. B {\bf 146}, 299 (1984);
                    D.V. Bugg {\it et al}.,
                    Phys. Lett. B {\bf 194}, 563 (1987).
\bibitem{Bru87}     W. Br\"uckner {\it et al}.,
                    Phys. Lett. B {\bf 197}, 463 (1987);
                    {\it ibid}. {\bf 199}, 596(E) (1987);
                    Zeit. Phys. {\bf A335}, 217 (1990).
\bibitem{Tim92}     R. Timmermans, Th.A. Rijken, and J.J. de Swart,
                    Nucl. Phys. {\bf A479}, 383c (1988);
                    Phys. Rev. D {\bf 45}, 2288 (1992).
\bibitem{Tab85}     F. Tabakin and R.A. Eisenstein, Phys. Rev. C {\bf 31},
                    1857 (1985).
\bibitem{Tim85}     P.H. Timmers, Ph.D. thesis, University of Nijmegen
                    (1985).
\bibitem{Nis85}     J.A. Niskanen, Helsinki preprint HU-TFT-85-28.
\bibitem{Koh86a}    M. Kohno and W. Weise, Phys. Lett. B {\bf 179}, 15 (1986);
                    Phys. Lett. B {\bf 206}, 584 (1988); Nucl. Phys. 
                    {\bf A479}, 433c (1988).
\bibitem{Fur87}     S. Furui and A. Faessler,
                    Nucl. Phys. {\bf A468}, 669 (1987).
\bibitem{Kro87}     P. Kroll and W. Schweiger,
                    Nucl. Phys. {\bf A474}, 608 (1987).
\bibitem{Alb88}     M.A. Alberg, E.M. Henley, and L. Wilets,
                    Phys. Rev. C {\bf 38}, 1506 (1988);
                    Z. Phys. {\bf A331}, 207 (1988).
\bibitem{LaF88}     P. LaFrance, B. Loiseau, and R. Vinh Mau, Phys. Lett.
                    B {\bf 214}, 317 (1988); P. LaFrance and B. Loiseau,
                    Nucl. Phys. {\bf A528}, 557 (1991).
\bibitem{Hai93}     J. Haidenbauer, K. Holinde, and J. Speth,
                    Nucl. Phys. {\bf A562}, 317 (1993).
\bibitem{Mae89}     P.M.M. Maessen, Th.A. Rijken, and J.J. de Swart,
                    Phys. Rev. C {\bf 40}, 2226 (1989).
\bibitem{Bra93}     F. Bradamante, R. Hess, and J.-M. Richard,
                    Few-Body Systems {\bf 14}, 37 (1993).
\bibitem{Sto93}     V. Stoks, R. Timmermans, and J.J. de Swart,
                    Phys. Rev. C {\bf 47}, 512 (1993).
\bibitem{Arn90}     R.A. Arndt, Z. Li, L.D. Roper, and R.L. Workman,
                    Phys. Rev. Lett. {\bf 65}, 157 (1990).
\bibitem{Bru86b}    W. Br\"uckner {\it et al}.,
                    Phys. Lett. B {\bf 169}, 302 (1986).
\bibitem{Bru85}     W. Br\"uckner {\it et al}.,
                    Phys. Lett. B {\bf 158}, 180 (1985).
\bibitem{Bru86a}    W. Br\"uckner {\it et al}.,
                    Phys. Lett. B {\bf 166}, 113 (1986);
                    Zeit. Phys. {\bf A335}, 217 (1990).
\bibitem{Lin87}     L. Linssen {\it et al}.,
                    Nucl. Phys. {\bf A469}, 726 (1987).
\bibitem{Hai92}     J. Haidenbauer, K. Holinde, V. Mull, and J. Speth,
                    Phys. Lett. B {\bf 291}, 223 (1992).
\bibitem{Alb93}     M.A. Alberg, E.M. Henley, L. Wilets, and P.D. Kunz,
                    Seattle preprint DOE/ER/40427-31-N93.
\end{thebibliography}
\end{document}